\documentclass[twocolumn,prl, showpacs,showpacs,aps]{revtex4}
\usepackage{graphicx}
\usepackage{dcolumn}
\usepackage{bm}
\usepackage{amsmath}
\usepackage{amssymb}
\usepackage{epsf}
\newcommand{\be}{\begin{equation}}
\newcommand{\ee}{\end{equation}}
\newcommand{\bea}{\begin{eqnarray}}
\newcommand{\eea}{\end{eqnarray}}
\newcommand{\mbf}{\mathbf}
\newcommand{\LD}{$\Lambda$-doublet~}

\begin{document}
\title{OH hyperfine ground state:  from precision measurement to molecular qubits}
\author{Benjamin L. Lev}
 \email{benlev@jila.colorado.edu}
\author{Edmund R. Meyer}
\author{Eric R. Hudson}\altaffiliation{Present address:  Department of Physics, Yale University, New Haven, CT 06520 USA}
\author{Brian C. Sawyer}
\author{John L. Bohn}
\author{Jun Ye}
\affiliation{JILA, National Institute of Standards and Technology and the University of Colorado \\ Department of Physics, University of Colorado, Boulder, Colorado 80309-0440, USA}
\date{\today}
\begin{abstract} 
We perform precision microwave spectroscopy---aided by Stark deceleration---to reveal the low magnetic field behavior of OH in its $^2\Pi_{3/2}$ ro-vibronic ground state, identifying two field-insensitive hyperfine transitions suitable as qubits and determining a differential Land\'{e} $g$-factor of $1.267(5)\times10^{-3}$ between opposite parity components of the $\Lambda$-doublet.  The data are successfully modeled with an effective hyperfine Zeeman Hamiltonian, which we use to make a tenfold improvement of the magnetically sensitive, astrophysically important $\Delta F=\pm1$ satellite-line frequencies, yielding $1\:720\:529\:887(10)$ Hz and $1\:612\:230\:825(15)$ Hz.

\end{abstract}
\maketitle
Scientific advances such as creating new forms of dipolar matter, realizing molecular scalable quantum computation, and uncovering physics beyond the Standard Model demand improved understanding of even the most familiar of polar molecules.  Techniques such as buffer gas cooling~\cite{Doyle98}, photoassociation~\cite{DeMille05}, and Stark deceleration~\cite{Meijer99,Bochinski03}, are beginning to yield cold samples of ground state polar molecules, such as OH, that can be utilized for these endeavors.  However, so far no technique simultaneously achieves the large molecule number and ultracold, $<$100 $\mu$K, temperature required.  Solving this problem would be exciting not only for the accompanying ability to study ultracold collisions and collective phenomena dominated by long-range dipole-dipole interactions~\cite{Bohn02}, but also for the prospect of using the strong dipolar interaction to robustly implement quantum logic gates~\cite{DeMille02,Zoller06}.  This would obviate the need for the precise, dynamical position control or the short-lived excited states that encumber other proposed quantum computation systems~\cite{DeMille02}.

Stark deceleration currently provides relatively large numbers of polar molecules, but at temperatures limited to a few mK.  Additional cooling will be required if we hope to reach densities necessary for exploring ultracold dipolar gas phenomena.  Magnetic trapping is a promising aid for cooling because it provides long term confinement of relatively hot particles while also admitting electric (E) fields to the trapped molecules.  This latter point is important since B- and E-fields can modify collisional processes in ways that enhance the efficiency of sympathetic cooling with co-trapped alkalis.  Once cooled, E-fields can tune the dipole interaction to explore collective phenomena such as field-linked states~\cite{Bohn02}.    To achieve these goals, we need to understand precisely the influence of B-fields on collisional resonances in the cold and ultracold regimes, and experimental data are required to validate theory models of polar molecular states in low B-fields.  Additionally, because fluctuating B-fields are a nuisance for long-term qubit coherence and precision measurements, accurate models are required to account for and nullify these unwanted effects.  The natural combination of precision microwave spectroscopy with the cold, monoenergetic packets from Stark decelerators~\cite{Hudson06} enables us to understand polar molecule hyperfine Zeeman behavior, as presented in this Letter.  
 
\begin{figure}[t]
\scalebox{0.44}[0.44]{\includegraphics{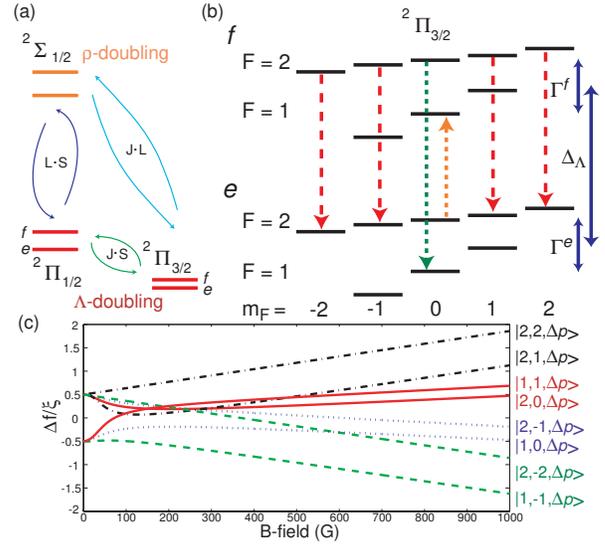}}
\caption{\label{fig1}(color online) (a) OH angular momentum perturbations. $\mbf{J}$, $\mbf{L}$, and $\mbf{S}$ are the total (minus nuclear), orbital, and spin angular momentum operators. (b) Small B-field Zeeman shifts in $^2\Pi_{3/2}$.  Long-dashed arrows depict main-line $\pi$-transitions for $\delta g_F$ measurement while short-dashed show measured satellite lines. (c) Main-line $\pi$-transition frequency shifts, $\Delta f$, versus B-field.  Curves of identical dash pattern asymptote to same set of $J$-basis lines. $\Delta_{\Lambda}$$\approx$$1.7$ GHz omitted from vertical scale.} 
\end{figure} 
As a demonstration of this technique, we experimentally verify our refinement of previous OH Zeeman theory~\cite{Townes55,Radford} to the hyperfine regime and use it to identify two candidate qubit transitions that are robust against B-field fluctuations.  Moreover, we apply this model to the precision measurement of the magnetically sensitive OH ground state satellite-lines, completing the ten-fold resolution improvement in all four lines of the astrophysically important $J=3/2$ \LD\cite{Hudson06}.  These measurements highlight the ability of the Stark decelerator not only to enhance our understanding of unexplored regimes of molecular coupling, but also to contribute toward searches of non-Standard Model physics, e.g.\ the variation of fundamental constants such as fine-structure, $\alpha$, and the electron-proton mass ratio, $y$, which may be measured by comparing Earth-bound OH with that found in OH megamasers~\cite{Darling03, Kanekar03}.  These distant sources are spatially well-defined, and by combining our recent measurements with astrophysical studies of comparable resolution, we will be able to constrain---with spatial dependence---fine-structure variation below 1 ppm for $\Delta\alpha/\alpha$ over 6.5 G yr~\cite{Kanekar05}.   We further note that the low-field behavior revealed in OH may also occur in other molecules and certainly in those with $^2\Pi$ structure.  Methods presented here are crucial if B-field effects are to be enhanced for molecular cooling schemes or mitigated in molecular-based clocks, qubits~\cite{DeMille02, Zoller06}, and precision measurements of the electron electric dipole moment in molecules~\cite{Kozlov95}.

The hyperfine Zeeman structure of OH is subtly perturbed by angular momentum couplings of higher order than the well-known, diagonal $\mbf{L}\cdot\mbf{S}$ spin-orbit interaction, and this study reveals their effect on OH transition frequencies in low B-field.  These perturbations cause the parity states, labeled $f$ and $e$, of the $^2\Pi_{3/2}$ ro-vibronic ground state---which are otherwise identical---to possess slightly different hyperfine energy splittings and g-factors.   Figure~\ref{fig1}(a) sketches the relevant energy levels and the higher-order angular momentum couplings between them.  The two $^2\Pi$ fine-structure manifolds are mixed with the first electronically excited state, $^2\Sigma_{1/2}$, via $\mbf{J}\cdot\mbf{L}$ and the off-diagonal $\mbf{L}\cdot\mbf{S}$ couplings.  Mediated by these couplings, the spin-rotation induced splitting of the $\rho$-doubled $^2\Sigma_{1/2}$ state causes the $^2\Pi$ states to also split.  The resulting \LD energy splitting, $\Delta_\Lambda$, between parity states is smaller in $^2\Pi_{3/2}$ due to the weaker, off-diagonal $\mbf{L}\cdot\mbf{S}$ coupling.   The hyperfine parity difference may be ascribed to the minute difference in coupling strength between $^2\Sigma_{1/2}$ and the energetically separated parity states.  Likewise, the larger $f$-state $g$-factor is due to the stronger $\mbf{J}\cdot\mbf{S}$ coupling between the energetically closer $^2\Pi$ $f$-parity states~\cite{Brown}.  

Experimentally, we focus on the electric dipole allowed, $\pi$ ($\Delta m_F$=$0$) and $\sigma$ ($\Delta m_F$=$\pm1$) transitions between the \LD parity states, as shown by arrows in Fig.~\ref{fig1}(b).  We write these states in the $|F,m_F,p\rangle$ basis and distinguish $\pi$-transitions from states by substituting $\Delta p$ for the parity value, $p$.  The hyperfine energies of the parity states are written as $\Gamma^{f,e}=\bar{\Gamma}\pm\xi/2$, where the average splitting, $\bar{\Gamma}$, and difference, $\xi$, are approximately 54.1 and 1.96 MHz, respectively~\cite{terMeulen72}.   Similarly, the g-factors are written $g^{f,e}=\bar{g}\pm\delta g/2$, where $\delta g$ is 0.3\% of $\bar{g}$~\cite{Radford}.  
 
\begin{figure}[t]
\scalebox{0.49}[0.49]{\includegraphics{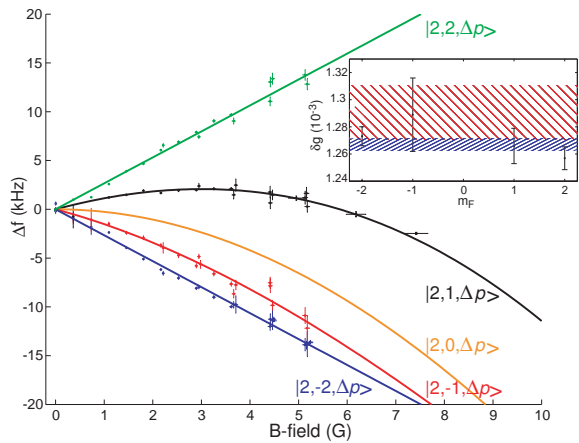}}
\caption{\label{fig2}(color online) Data and theory fits for energy differences between main-line 
  $|2,m_F,\Delta p\rangle$ 
$\pi$-transitions versus B-field. Inset shows summary of $\delta g$ data.  Red (large) hatch is previous $\delta g_J$ data~\cite{Radford}, and blue (small) hatch is $\overline{\delta g_F}$.}
\end{figure}

This work distinguishes itself from earlier studies~\cite{Radford, Vanderlinde80} in which relatively poor spectroscopic resolution of the ground state precluded low-field measurements.  While these subtle differences can be neglected in previous experiments wherein huge magnetic fields cause the Zeeman energies to be much larger than the hyperfine splittings, we demonstrate that---in this new era of high-resolution molecular spectroscopy---these effects cannot be ignored when low-fields are present.  In other words, in fields of 1 T, where $J$ is a good quantum number, \LD transition frequency differences vary as expected---linearly with applied B-field~\cite{Radford}.  However, in fields $<$30 G, where $F=J\pm1/2$ is a good quantum number, we do not find a linear variation of transition frequency with B-field.  We instead observe a linear frequency shift out to $\sim$1 G, and thereafter detect shifts that are significantly varied in curvature versus B-field, as shown in the data of Fig.~\ref{fig2} for $\mbox{B}<8$ G. 

Qualitatively, this effect may be understood as a differential onset of the Paschen-Back effect in the two parity states, which we explicate by first considering the hypothetical case in which no differences exist between either the $g$-factors ($\delta g=0$) or the hyperfine splittings ($\xi$ = 0).  As such, the parity states are identical with respect to an applied B-field, and one would expect no frequency shifts in the hyperfine Zeeman transitions of, e.g., the $\pi$-transitions between main-lines ($\Delta F=0$) versus B-field until $\mbox{B}\gg1$ T.  This holds in spite of quadratic level shifts induced by mixing between $F=2$ and $F=1$ at $B>30$ G, as described by the Paschen-Back effect, since this onset occurs at exactly the same B-field for $p=f,e$.  If we now account for $\delta g\neq0$, then a linear variation of the hyperfine Zeeman transitions appears and is proportional to $m_F\delta g$ at low fields.  Finally adding the hyperfine perturbation induces an interesting dynamic in the frequency differences.  The onset of the Paschen-Back effect now occurs at slightly different B-fields for the parity states, causing the otherwise linear differential frequency shifts to become quadratic at small B-fields for transitions involving $|m_F|=1$.

To investigate this more rigorously, we employ an effective Hamiltonian~\cite{Brown} to systematically distill the full---though intractable---system Hamiltonian into one that focuses solely on the state of interest, $^2\Pi_{3/2}$.  The resulting Zeeman Hamilitonian is expanded into the hyperfine basis, and the hyperfine energy and $g$-factor differences are incorporated parametrically, a point of departure from previous treatments~\cite{Brown,Townes55}.  The resulting hyperfine Hamiltonian is applicable to higher-rotational states ($J\geq3/2$) as well as to the $^2\Pi_{1/2}$ manifold~\footnote{Rotation and nuclear $g$-factors modify $\delta g$ by $\leq$0.1\%.}:  \bea\label{eq:H} \begin{array}{l}H^{f,e}_{^2\Pi_{3/2}}= \\ \begin{array}{c}F=J+1/2\ \ \ \ \ \ \ \ \ \ F=J-1/2\\\left(\begin{array}{cc}\frac{\Gamma^{f,e}}{2}+2\beta^{f,e} m_F \frac{J}{2J+1} & \frac{ \beta^{f,e}\sqrt{(J+\frac{1}{2})^2-m_F^2}}{2J+1} \\ &  \\\frac{\beta^{f,e}\sqrt{(J+\frac{1}{2})^2-m_F^2}}{2J+1} &    -\frac{\Gamma^{f,e}}{2}+2\beta^{f,e} m_F \frac{J+1}{2J+1} \end{array}\right),\end{array}\end{array}\eea where $\beta^{f,e}=\mu_0 Bg^{f,e}$.  Equation~\ref{eq:H} represents two parity Hamiltonians for each value of $m_F$, and upon diagonalization yields 16 energy eigenvalues for the four $J=3/2$ hyperfine states.  The off-diagonal terms vanish for $|m_F|=2$ as expected for states which do not couple to $F=1$, and the diagonal Zeeman terms vanish for the linearly insensitive $m_F=0$ states.  Eigenvalue differences correspond to transition energies as  a function of B-field.  

The main-line $\pi$-transitions used to investigate this model are depicted in Fig.~\ref{fig1}(b) and plotted in Fig.~\ref{fig1}(c) as a function of B-field using the eigenvalues from Eq.~\ref{eq:H}.  All lines---except those with $|m_F|=2$---significantly depart from linearity before they asymptote to their $J$-basis behavior.  Of note is the $|2,1,\Delta p\rangle$ line which exhibits a negative frequency shift from its zero field value over a region spanning nearly 500 G.  Figure~\ref{fig2} displays the $F=2$ $\pi$-transitions measured in the low-field, $\mbox{B}<10$ G regime.  Evidently, the $|2,1,\Delta p\rangle$ line crosses zero frequency shift twice before pairing with its $J$-basis partner, $|2,2,\Delta p\rangle$, at high B-field.  Interplay between two small quantities---$\xi$ and $\delta g$---governs this behavior, which may be seen from an expansion of the eigenvalue energy differences to third-order in B-field, valid up to $\sim$30 G:  \bea\label{energies} \begin{array}{r}\Delta E_{\Delta F=0}=\frac{3m_F\delta g}{2J+1}\mu_0B+\frac{(\bar{g}_J\mu_0B)^2}{\bar{\Gamma}}\frac{((J+\frac{1}{2})^2-m_F^2}{4(J+\frac{1}{2})^2}\ \times\\\left[\frac{2\delta g}{\bar{g_J}}-\frac{\xi}{\bar{\Gamma}}+\mu_0B\left(\frac{3\delta g}{\bar{\Gamma}}-\frac{2\bar{g}_J\xi}{\bar{\Gamma}^2}\right)\frac{m_F}{J+\frac{1}{2}}\right].\end{array}\eea  A practical consequence of this interplay is the insensitivity to magnetic fluctuations of the $|2,1,\Delta p\rangle$ and $|1,-1,\Delta p\rangle$ lines, indicated in Figs.~\ref{fig2} and~\ref{fig1}(c), at B = 3.01(2) G and 7.53(4) G, respectively.  With experimentally achievable 1 mG B-field stability, we estimate frequency fluctuations to be $<$5 mHz, which provides a transition stability superior to that of the $m_F=0$ satellite-line components in zero field.  In addition, each is a transition between states of equal, non-zero magnetic moment, allowing stable magnetic trapping without heating.  These states have large, equal-but-opposite electric dipoles ($|\mu|=1.67$ D) which are easily aligned with low E-fields ($\sim$1 kV/cm), and as such are enticing candidates for encoding B-field stable molecular qubits~\cite{DeMille02}. 

\begin{figure}[t]
\scalebox{0.5}[0.5]{\includegraphics{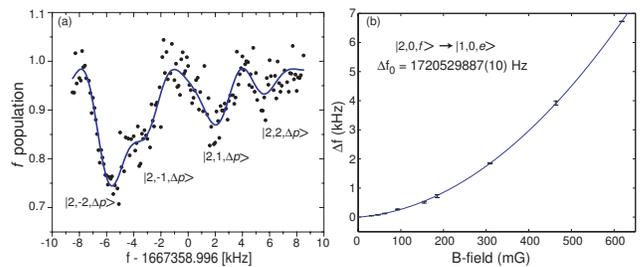}}
\caption{\label{fig3} (a) Rabi spectroscoy of the $|2,m_F,\Delta p\rangle$ main line at $\mbox{B}=2.96$ G. (b)  Fit to $|2,0,f\rangle\rightarrow|1,0,e\rangle$ satellite line.  Each data point is the weighted average of several measurements taken over many days.}
\end{figure} 
To experimentally validate our model, we utilize the previously discussed Stark decelerator and microwave spectroscopy~\cite{Hudson06} to obtain high spectral resolutions in the presence of $<$10 G magnetic fields.  In brief, the Stark decelerator provides OH primarily in the strongest, weak electric (E) field-seeking states, $|F,m_F,p\rangle=|2,\neq$$0,f\rangle$.  Typical operation produces a $10^6$ cm$^{-3}$ dense packet with mean velocity controllable from 410 m/s down to nearly rest and a minimum temperature of 5 mK~\cite{Bochinski03}.   We choose to run the decelerator at 200 m/s as a compromise between slowing efficiency, microwave power, and interrogation time~\cite{Hudson06}.   We perform Rabi spectroscopy by interrogating the OH packet with one or more microwave pulses referenced to the Cs standard~\cite{Hudson06}.  Laser-induced fluorescence provides state detection by exciting the $f$ manifold---equally driving the two hyperfine states---with 282 nm light and registering a fraction of the subsequent 313 nm emission with a photomultiplier tube.  Main and satellite transitions originating from $|2,m_F,f\rangle$ are driven with a single microwave pulse and detected by registering a reduction in photon counts (see Fig~\ref{fig1}(b)).

The 10 cm long, TM$_{010}$ microwave cavity is tuned near the $|2,m_F,\Delta p\rangle$ main line.  A solenoid is wound around the cavity and encased in a $\mu$-metal magnetic shield.  The latter reduces the ambient field to $\leq$6 mG, while the former can apply up to a 10 G field along the resonant TM$_{010}$ axis, which itself is aligned along the OH beam path.  At the $|2,m_F,\Delta p\rangle$ frequency, the cavity E- and B-fields are nearly collinear---determined by the absence of observed $\sigma$-transitions---and the E-field magnitude is constant over 80\% of the cavity.  In contrast, the satellite-line frequencies are detuned far enough from cavity resonance to distort the associated intracavity E-field magnitude and direction, and we observe both $\pi$- ($\mbf{E}\parallel\mbf{B}$) and $\sigma$- ($\mbf{E}\perp\mbf{B}$) transitions on these lines.

We calibrate the intracavity B-field by tracking the $|2,m_F,f\rangle\rightarrow|1,m'_F,e\rangle$ satellite splitting as a function of B-field in the $<$1 G regime where the Zeeman theory is well-understood.   Concurrent observation of the $\pi$- and $\sigma$-lines provides cross-checks on the B-field calibration.  Their relative peak heights and frequency shifts versus cavity position also provide a tool for mapping E-field direction and power inhomogeneities.  When combined with the $|2,m_F,\Delta p\rangle$ data, this mapping provides a measurement of the B-field inhomogeneity as well.  We perform B-field calibration and corresponding Zeeman shift measurements over several portions of the cavity.  This reduces systematic errors, and we conclude that our B-field calibration and inhomogeneity mapping are accurate to within less than 1\%. Other systematic shifts, such as from collisions, stray electric field, doppler shift, and black body radiation, are collectively $<$3 Hz~\cite{Hudson06}.

Figure~\ref{fig3}(a) shows typical main-line data.  Only four peaks are visible since for $\pi$-transitions $m_F=m'_F=0$ is selection-rule forbidden when $\Delta F=0$.  The data are fit with four Rabi peaks of the form $\sin^2(\frac{\omega'_Rt}{2})$, where $\omega'_R=\sqrt{\omega^2_R+\delta^2}$, and $\omega_R$ and $\delta$ are the Rabi frequency and detuning, respectively.  Each line's detuning is a free parameter.  Figure~\ref{fig2} summarizes results taken at different B-fields.  The larger error bars are from traces taken at microwave interrogation times as short as $t=70$ $\mu$s, which have larger transit-limited resolutions than those taken at $400$ $\mu$s.   As expected, the $m_F=m'_F=\pm 2$ lines are linear and we detect curvature in the $m_F=m'_F=\pm1$ lines at fields greater than $\sim$1 G.  The theory and data show excellent agreement without any free parameters.  The only parameter in Eq.~\ref{eq:H} with uncertainty greater than $\sim$$10^{-5}$ is $\delta g$, which was measured by Radford in the $J$-basis at B = 0.6-0.9 T to be $\delta g_J=1.29(2)\times10^{-3}$.  Coupling to higher rotational states at large fields may inflate the $g_J$ difference over that measured in the $F$-basis regime.  To test this, we perform a single-parameter fit to our low-field data to obtain $\delta g_F$.  The resulting fits are shown in Fig.~\ref{fig2}, which are indistinguishable from the theory curves. The inset plots the measured $\delta g_F$ for each of the four $m_F$ states, along with their weighted mean, $\bar{\delta g}_F=1.267(5)\times10^{-3}$. This value is $1\sigma$ consistent with Radford's $g_J$, but higher precision measurements in both regimes may distinguish a difference.

As a further application of this hyperfine Zeeman theory for OH, we measure the magnetically sensitive satellite-line transition frequencies.  Specifically, we trace the $\nu_{21}=|2,0,f\rangle\rightarrow|1,0,e\rangle$ and $\nu_{12}=|2,0,e\rangle\rightarrow|1,0,f\rangle$ frequencies versus B-field.  These Zeeman transitions are first-order insensitive to B-field, which suppresses spurious frequency shift contributions from calibration and ambient field uncertainties.  Figure~\ref{fig3}(b) shows the approximately quadratic variance of $\nu_{21}$ versus applied B-field.  Nearly overlapping $\pi$- and $\sigma$-transitions preclude an accurate measurement at zero applied field, and instead we must magnetically split the lines. The theory fit is used to project the data down to zero applied B-field where the ambient field is the only magnetic perturbation.  The uncertainty due to the ambient field on the line frequencies is estimated to be $<$0.6 Hz.  The fit value for the B-field calibration is consistent with that obtained earlier, albeit with a larger uncertainty.  We choose to use the fit-value's error to obtain a conservative estimate of the transition measurement's uncertainty.  The fit has a reduced $\chi^2=1.2$, and projecting the curve to zero applied field results in a bare frequency of $\nu_{21}$ = $1\:720\:529\:887(10)$ Hz.  For the $\nu_{12}$ line, we use the distorted E-field at the cavity entrance to drive $|2,\pm1,f\rangle$$\rightarrow$$|2,0,e\rangle$ $\sigma$-transitions.  A second microwave pulse then drives the $\pi$-transition to $|1,0,f\rangle$.  Using the previous fit's calibrations, we measure $\nu_{12}$ = $1\:612\:230\:825(15)$ Hz as the transition's bare frequency.  These values provide a tenfold improvement over the previous measurement accuracy of 100-200 Hz~\cite{terMeulen72}.  

Variation of $\alpha$ may be constrained by taking the sum and difference of either the main- or satellite-lines.   The difference is sensitive to $\alpha$ while the sum measures the red-shift systematic.  The satellite-lines offer a 55-fold increase in resolution over the main-lines due to the $2\bar{\Gamma}$ versus $\xi$ separation~\cite{Darling03}.  Now that the complete \LD is measured with high-precision, one can simultaneously constrain variation of $\alpha$ and $y$~\cite{Kanekar03}. All four lines must be detected from the same source, and the closure criterion, i.e., the zero difference of the averages of the main- and satellite-line frequencies, provides a critical systematics check.  We obtain an Earth-bound closure of 44$\pm$21 Hz.  Imminent astrophysical measurements of the satellite-lines below the 100 Hz level~\cite{Kanekar06} highlight the prescience of our study for observing the variation of fundamental constants.  

We thank DOE, NRC, NSF, and NIST for support.  

\end{document}